\documentclass[runningheads]{llncs}
\usepackage[export]{adjustbox}[2011/08/13]
\usepackage[numbers]{natbib}
\usepackage{graphicx}
\usepackage{tabularx} 
\usepackage{float}
\usepackage{algorithm2e}
\usepackage{xcolor}

\usepackage[bottom]{footmisc}
\begin{document}
%

\title{Towards Agent-based Models of Rumours in Organizations: A Social Practice Theory Approach\thanks{This paper has been peer-reviewed and accepted for the Social Simulation Conference 2018 in Stockholm. The final authenticated version will be available online at Springer LNCS. The DOI will be provided when available.}}
\titlerunning{Rumourmongering in Organizations}
%
\author{Amir Ebrahimi Fard \and
Rijk Mercuur\and
Virginia Dignum \and
Catholijn M. Jonker \and
Bartel van de Walle}
\authorrunning{A. Fard et al.}
%
\institute{Delft University of Technology \\
\email{A.EbrahimiFard@tudelft.nl},
\email{R.A.Mercuur@tudelft.nl},
\email{M.V.Dignum@tudelft.nl},
\email{C.M.Jonker@tudelft.nl} and
\email{B.A.vandeWalle@tudelft.nl }}
\maketitle              

\begin{abstract}
Rumour is a collective emergent phenomenon with a potential for provoking a crisis. Modelling approaches have been deployed since five decades ago; however, the focus was mostly on epidemic behaviour of the rumours which does not take into account the differences of the agents. We use social practice theory to model agent decision making in organizational rumourmongering. Such an approach provides us with an opportunity to model rumourmongering agents with a layer of cognitive realism and study the impacts of various intervention strategies for prevention and control of rumours in organizations.
\end{abstract}

\section{Introduction}
The phenomenon of rumourmongering has malicious impacts on societies.
Rumours make people nervous, create stress, shake financial markets and disrupt aid operations \cite{Vosoughi2018}. In organizations, rumours lead to unpleasant consequences such as, breaking the workplace harmony, reduction of profit, drain of productivity and damaging the reputation of a company \cite{DiFonzo1994,Michelson2000}. Recent work on the McDonald's wormburger rumour and the P\&G Satan rumour confirm the negative impact of rumours on the productivity of firms \cite{DiFonzo1994}.

For 120 years, scholars from a wide range of disciplines are trying to understand different dimensions of this phenomenon. Research in rumour studies can be classified according to the approach followed: a case-based approach and a model-based approach. In the case-based approach, results are based on case studies, not on models, making it hard to generalize their conclusions. The model-based approach tries to explain the phenomenon of rumours by model-based based simulations. The model-based approaches, so far, focus only on the dynamic of the spread, while rumour is a collective phenomenon and the acts of individuals can influence the whole system. Rumours in organizations have been mainly approached with case-based studied and dynamic spreading model. To our knowledge there are no studies where the cognition of the individual is taken into account.

In our agent-based approach, we study the dynamics of the spread of rumours in organizations as an emergent (collective) behavior resulting from the behavior of individual agents using social practice theory. We use the proposed model to study the impact of change in organizational layout on control of organizational rumour.

The concept of social practices stems from sociology, and aims to depict our `doings and sayings' \cite[p.~86]{Schatzki1996}, such as dining, commuting and rumourmongering. This paper uses the semantics of the social practice agent (SoPrA) model \citep{MercuurSoPrA} to gain insights in rumourmongering in organizations.\footnote{\citet{MercuurSoPrA} provides a static model of SoPrA based on literature and argued modelling choices. This paper applies this model to the domain and extends it by including competences and affordances and modelling a dynamic component based on \citep{Mercuur2015}. Note that \citet{MercuurSoPrA} is still under review and only available as pre-print at the moment of writing.} SoPrA provides an unique tool to combine habitual behavior, social intelligence and interconnected practices in one model. This makes SoPrA especially well-suited for studying the spread of rumours in organizations as this practice is largely habitual, social \citep{Hackman1990} and interconnected with practices as working and moving around. To build the model with SoPrA, owing to lack of available empirical dataset, we give a proof-of-concept on how to collect data by doing eight semi-structured interviews.  

This paper is organized as follows. The next section provides an overview of the research on rumours with an emphasis on studies of organizational rumour. Section~\ref{context} describes the context for our experiment, and the methodology of data collection and data preprocessing. The model is introduced in Section~\ref{modelsection}. One possible experiment is described in Section~\ref{experiments} and Section~\ref{concl} presents our conclusions, discussion and ideas for future work.

\section{Background \& Related Work}
\label{sec:literature}
Rumours are unverified propositions or allegations which are not accompanied by corroborative evidence \cite{DiFonzo1994}. Rumours take different forms such as exaggerations, fabrications, explanations \cite{Prasad1934}, wishes and fears \cite{Knapp1944}. Rumours have a lifecycle and change over the time. Allport and Postman in their seminal work “psychology of rumour” concluded that, “as a rumour travels, it grows shorter, more concise, more easily grasped and told.” \cite{Allport1965}. Buckner considers rumour a collective behaviour which is becoming more or less accurate while being passed on as they are subjected to the individuals' interpretations which depends on the structure of the situation in which the rumour originates and spreads subsequently \cite{Buckner1965}. Rumours are conceived to be unpleasant phenomena that should be curtailed. Therefore, a number of strategies have been proposed to prevent and control them \cite{Buckner1965,Oh2013, Knopf1975}.

One of the rumour contexts that has received attention from researchers for almost four decades is organizations. Like rumour in general context which is explained in above paragraph, rumour in organizations has different types and follow its own life-cycle \cite{DiFonzo1994, DiFonzo1998, Bordia2003, Kimmel2004, Bordia2006, Bordia2014}. Also, to quell credible and non-credible organizational rumours, a number of different techniques and strategies have been suggested \cite{DiFonzo1994,Kimmel2004}. The research approach also follow the same pattern, with a slight difference which to best of our knowledge is qualitative without adopting any modelling approach.

The related literature reported above are based on case studies or experiments in the wild. This pertains to the types of rumour, dynamics of rumour and strategies to control rumours, either in general or in organizational contexts. These case studies and experiments are to inform the construction of theories and models underlying the phenomenon of rumourmongering. Theories and models, in turn, should be tested in case studies and simulations. Model-based approaches do just that. However, the current state-of-the-art in model-based simulations of rumourmongering focus only on the dynamics of the rumourmongering, comparable to the epidemic modelling and spread of viruses \cite{Daley1964, Zanette2002, Nekovee2007, Wang2017, Turenne2018}. These models do not consider the complexities of the agents that participate in rumourmongering.

The research area of agent-based social simulations (ABSS) specializes on simulating the social phenomena as phenomena that emerge from the behaviour of individual agents. ABSS is a powerful tool for empirical research. It offers a natural environment for the study of connectionist phenomena in social science. This approach permits one to study how individual behaviour give rise to macroscopic phenomenon \cite{Epstein1999}. Such an approach is an ideal way to study the macro effects of various social practices, because it can capture routines which are practiced by individuals on a regular basis in micro level and see their collective influence in a macro level.  

\section{Domain}\label{context}
This research investigates the daily routine of rumourmongering in a faculty building on the campus of a Dutch University. In this faculty, students, researchers and staff work in offices with capacity of one to ten people. 
Aside from the actual work going on in the building, filling a bottle with water, getting coffee from the coffee machine, having lunch at the canteen and going to the toilet are among the most obvious practices that every employee in this faculty does on a daily basis.

Nevertheless, there are other daily routines in the organization which are not that obvious. One of these latent routines is rumourmongering. Rumours or unverified information are transferred between students, researchers and staffs on a daily basis, during lunch, while queuing for coffee, when seeing each other in the hallways, and when meeting in classrooms and offices. All these situations are potential contexts for casual talks and information communication without solid evidence.
 
For data collection we conducted semi-structured explorative interviews with people from the above-mentioned faculty. Semi-structured interviews allows us to ask questions that are specifically aimed at acquiring the content needed for the SoPrA model, while still giving the freedom to ask follow up questions on unclear answers. The data collection can be improved in future works by increasing the number of interviewees and diversifying them (Not only asking from students).
For demographic information, the reader is referred to  Table~\ref{demographics}. We prepared following question set to ask from each interviewee based on the meta-model which will be explained in the next section:

\begin{enumerate}
\item What are the essential competencies for rumourmongering?
\item What are the associated values with rumourmongering?
\item What kind of physical setting is associated with rumourmongering?
\end{enumerate}

\begin{table}[ht]\label{demographics}
\footnotesize
\centering
\caption{Interviewees demography.}
\label{table:demography}
\begin{tabularx}{\columnwidth}{|X|X|l|X|X|}
\hline
\textbf{Number of Interviews} & \textbf{Number of Different Countries} & \textbf{Lowest Educational Level} & \textbf{Mean age} & \textbf{Female \%}
\\ \hline
8 & 6 & MSc & 28 & 50
\\ \hline
\end{tabularx}
\end{table}

Given the thin line between personality traits and competences, we used the Big Five model \cite{Goldberg1993} to differentiate between personality traits and competences. For Question 2, we asked the interviewees to choose the relevant values from Schwartz's Basic Human Values model \cite{Schwartz2012}. We asked the same set of questions about fact-based talk.

We processed the collected data in two ways before using it in the model. Firstly, we clustered answers that point to the same concept. For example, in Question 3, interviewees gave answers such as cafeteria, coffee shop and cafe to point to a place where people can get together and drink coffee. In the coffee example, we clustered answers under the term of "coffee place".

Secondly, we classified the answers to Question 2. As mentioned, for that question, we asked interviewees to pick associated values from Schwartz's Basic Human Values model. We used the third abstraction level of the model which is more fine-grained and compared to other levels, and gave the interviewees a better idea of what they point to. However, a model based on level three, would not allow us to compare the agents effectively. Therefore, we decided to wrap the answers and classify them based on second abstraction level. Using a  classification based on the first abstraction level would have been too homogeneous in the sense that the agents would behave too similar, which would loose the effectiveness of the simulation.

\section{Model}
\label{modelsection}
The model has two main parts: (i) static part and (ii) dynamic part. In the static part, the components of the model and their properties are described, and in the dynamic part we explain the interaction of those components. 

\subsection{Static Part}
This section describes the SoPrA meta-model which is used as the groundwork for our agent-based model, how we use empirical data to initiate the model, the model choices we make and how we tailor the model to the context of organization.

\begin{figure}[ht]
\includegraphics[height=0.4\textheight, center]{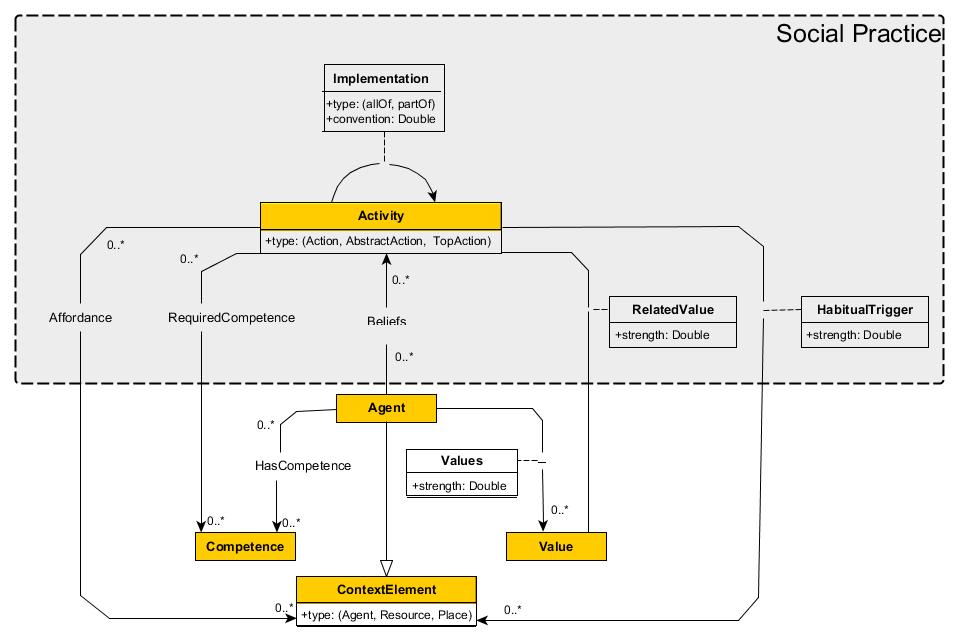}
\caption{The social practice meta-model captured in the Unified Modelling Language, including classes (yellow boxes), associations (lines), association classes (transparent boxes), navigability (arrow-ends) and multiplicity (numbers).}
\label{fig:uml}
\end{figure}

The SoPrA meta-model was introduced by \citet{MercuurSoPrA} and describes how the macro concept of social practices can be connected to micro level agent concepts. Figure \ref{fig:uml} shows SoPrA in a UML-diagram. The main objects in a SoPrA model are activities (e.g., fact talk, rumourmongering), agents (e.g., PhD students, supervisors), competences (e.g., networking, listening), context elements (e.g., office, cafetaria) and values. Values here refer to human values as found by the earlier stated Schwartz model, such as, power or conformity. The social practice is an interconnection of (1) activities and (2) related associations as depicted by the grey box in Figure \ref{fig:uml}. For example, the practice of talking consists of two possible activities fact talk or rumourmongering. The social practice connects these different activities with the \texttt{Implementation} association. If activity $A$ implements activity $B$ this means that $A$ is a way of or a part of doing $B$.

The \texttt{Implementation} association is the first of several associations that are related to an activity (see Table  \ref{table:assocations}).  Most associations are fairly self-explanatory, however the \texttt{Trigger} and \texttt{Strategy} association are a bit more complex. Following \citet{Wood2007}, triggers are the basis for habitual behaviour. If an agent is near a context element that has a trigger association with an activity, then it will do that activity automatically (without for example considering its values). Following \citet{Ostrom2007}, strategies are related to norms and signify that something is the normal way to do something.. If an agent believes that activity $A$ is a strategy for activity $B$, then it believes that other agents usually implement activity $B$ by doing activity $A$.

The SoPrA meta-model does not only relate the activities to other classes, but the agent itself also has two types of associations: \texttt{HasCompetence} and \texttt{ValueAdherence} which plays a role in choosing the activities it will do:. The \texttt{HasCompetence} association links possible skills to the agent who masters those. The \texttt{ValueAdherence} association captures if an agent finds that value important.

\begin{table}[ht]
\footnotesize
\centering
\caption{The associations attached to the activity and their specification.}
\label{table:assocations}
\begin{tabularx}{\columnwidth}{|l|X|}
\hline
\textbf{Association}   & \textbf{Specification}                                 \\ \hline
Implementation     & which activities are a way of or a part of doing the activity      \\
Affordance         & which context elements are needed to do the activity                       \\
RequiredCompetence & which competences are needed to do the activity                           \\
Knowledge               & which activities an agent knows about                  \\
Belief             & which personal beliefs an agent has about the activity \\
RelatedValue       & which values are promoted or demoted by the activity                  \\
Trigger             & which context elements habitually start the activity   \\
Strategy          & which activities usually implement the activity        \\ \hline
\end{tabularx}
\end{table}

The model can be initiated using empirical data. Note that in this study we focussed on a small set of explorative interviews. We show with this initial data a proof-of-concept of how the model can be initiated. To properly ground the model a larger and more rigorous empirical study is necessary. 

\begin{figure}[h]
\includegraphics[width=0.9\paperwidth, center]{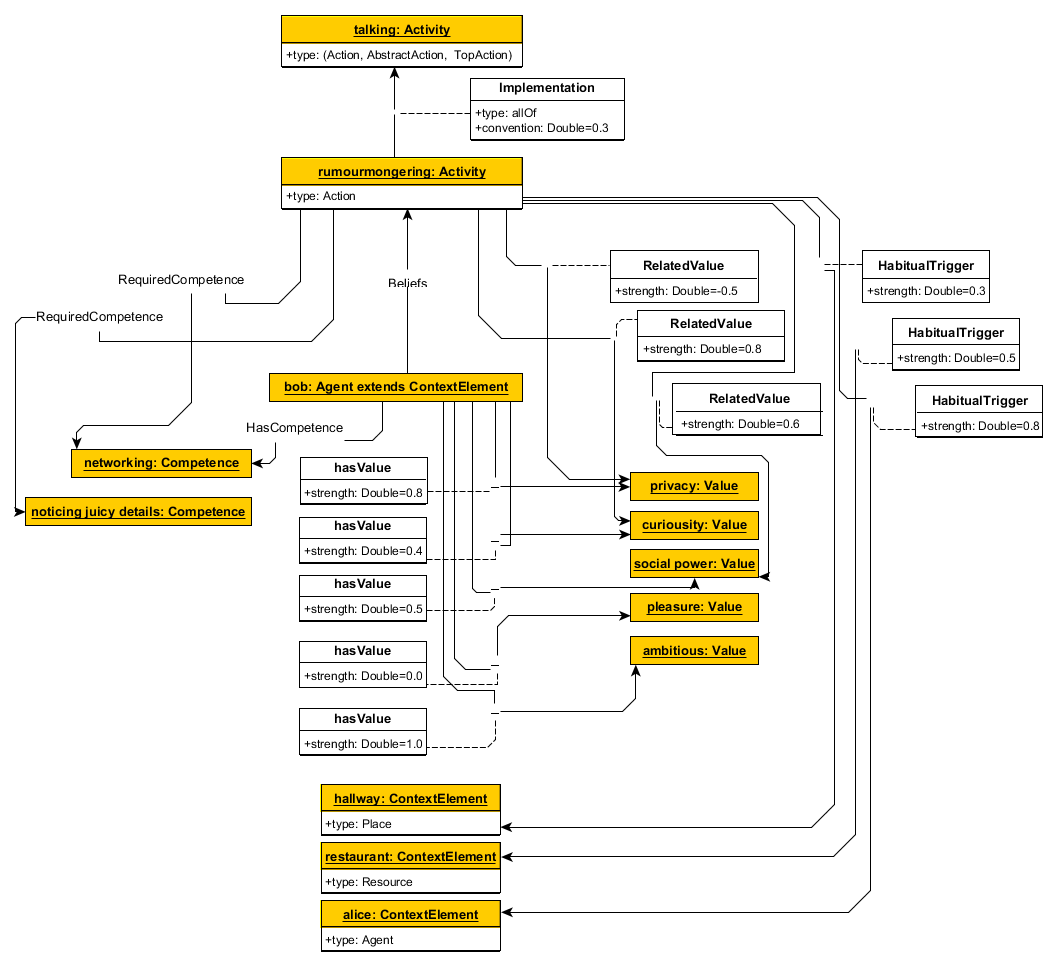}
\caption{An instance of the SoPrA meta-model for the activity of rumourmongering and one agent. For illustration purposes the assocations related to the activity 'talking and the agent 'alice' are omitted.}
\label{fig:umlinstance}
\end{figure}

The activity class has three instances: talking, rumourmongering and fact talk. The number of instances of agent can vary in the different experiments (see Section \ref{experiments}). The instances of the context element, competence and values class are based on the gathered data and can be found in Table \ref{table:rumourelements} and \ref{table:facttalkelements}.\footnote{The context-element `Friend' and `Colleague' are special cases; these are rather attributes of context-elements (i.e., agents) than context-elements themselves. In our model these are to some extent implicitly captured, because the agents who one sees most often (i.e., friends, colleagues) are mostly likely to be habitually associated with an action.} The complete static model consists both of object instances and associations between these instances. An example focusing on one agent (i.e., \texttt{Bob}) and one activity (i.e., \texttt{rumourmongering}) is shown in Figure \ref{fig:umlinstance}. Bob beliefs that the activity of rumourmongering is related to the value of privacy, curiosity and social power. He thinks it requires the competence of networking and noticing juicy details and thinks the activity is triggered (to some extent) by the hallway, restaurant and another agent named Alice. Furthermore, he himself has the competence of networking and adheres strongest to the value of ambition and weakest to the value of pleasure.

\begin{table}[H]
\caption{The elements associated with the rumourmongering activity.}
\label{table:rumourelements}
\begin{center}
\begin{tabular}{| c | c | c |}
\hline
\multicolumn{3}{| c |}{\textbf{Rumourmongering}} \\ 
\cline{1-3}
\textbf{Context Elements} & \textbf{Meaning} & \textbf{Competence} \\
\hline

Friend & Self-Direction &Sneaky Skills \\ \hline
Coffee place & Power & Network Skills \\ \hline
Hallway & Hedonism & Talking Skills \\ \hline
Restaurant & Achievement & Observing Skills \\ \hline
Office & Benevolence & \\ \hline
Phone &  &  \\ \hline
Computer &  &  \\ \hline

\end{tabular}
\end{center}
\end{table}

\begin{table}[H]
\caption{The elements associated with the fact talk activity.}
\label{table:facttalkelements}
\begin{center}
\begin{tabular}{| c | c | c |}
\hline
\multicolumn{3}{| c |}{\textbf{Fact talk}} \\ 
\cline{1-3}
\textbf{Context Elements} & \textbf{Meaning} & \textbf{Competence} \\
\hline

Colleague & Universalism & Being knowledgeable \\ \hline
Academic Staff & Self-Direction & Listening Skills \\ \hline
Office & Benevolence & Critical Thinking Skills \\ \hline
Conference & Achievement & Communication Skills \\ \hline
Meeting room & Tradition &  \\ \hline
Classroom &  &  \\ \hline
Restaurant &  &  \\ \hline
Phone &  &  \\ \hline
Computer &  &  \\ \hline
Pen &  &  \\ \hline
Coffee  &  &  \\ \hline

\end{tabular}
\end{center}
\end{table}

The agents differ in which activity they associate with which element. In other words, the SoPrA meta-model does not initiate one social practice that all agents share, but one social practice \textit{per agent}. The chance that an agent relates an activity to a competence is based on the empirical data we gathered in the interviews. For example, if 50\% of the interviewees linked critical thinking skills to fact talk the chance an agent makes this association depends on a binomial distribution with $p=0.5$. For \texttt{relatedValue} association and \texttt{HabitualTrigger} association all agents make the associations as mentioned in Table \ref{table:rumourelements} and \ref{table:facttalkelements}. However, the weights differ per agent. The weights for the \texttt{relatedValue} association are picked from a normal distribution between $0$ and $1$. Given the lack of empirical data on the relation between activities and human values, we follow the related finding of the World Value Survey that people adhere to values with roughly a normal distribution \citep{worldvaluesurvey}. The weights for \texttt{HabitualTrigger} are picked on a logarithmic distribution based on the empirical work of \citep{Lally2010}. One interesting modelling choice we made was to drop the \texttt{Affordance} assocations in the conceptual model. The SoPrA meta-model conceptualizes two associations with context elements. The \texttt{HabitualTrigger} association representing that some context element can automatically lead to a reactive action and the \texttt{Affordance} association representing that some context elements are a pre-condition to enact a certain behaviour. None of our interviewees mentioned a possible context element that affords rumourmongering fact talk. As such this association seemed irrelevant for our model.

The associations related to the agents themselves are based on random distributions. Each competence has a 50\% chance to be related to an agent. Each value is associated to each agent, but the weights differ. The weights for the \texttt{hasValue} association strength is based on a correlated normal distribution. \citet{Schwartz2012} shows that the strength to which people adhere to values is correlated. For example, people who positively value universalism usually negatively value achievement. We use the correlations found by \citet{Schwartz2012} to simulate intercorrelated normal distribution from which we pick the weights. In future work, we aim to extent our interviews to also gather data that can inform these weights. 

For our modelling context, we need to extend the SoPrA model with a spatial component. We do this by adding two attributes to the \texttt{ContextElement} class called \texttt{x-coordinate} and \texttt{y-coordinate}. These coordinates can be used by the agent to sense which objects are near. Note that every agent is also a context element as indicated with the 'generalization' association in the UML-diagram.

\subsection{Dynamic Part}
This section describes the dynamic part of the model which on each tick comprises:

\begin{enumerate}
\item An agent decides on its location using the moving submodel and updates its coordinate attributes.
\item An agent decides if it will engage in fact talk or rumourmongering based on the choose-activity submodel.
\end{enumerate}
 
\begin{figure}[H]
\includegraphics[width=\linewidth, center]{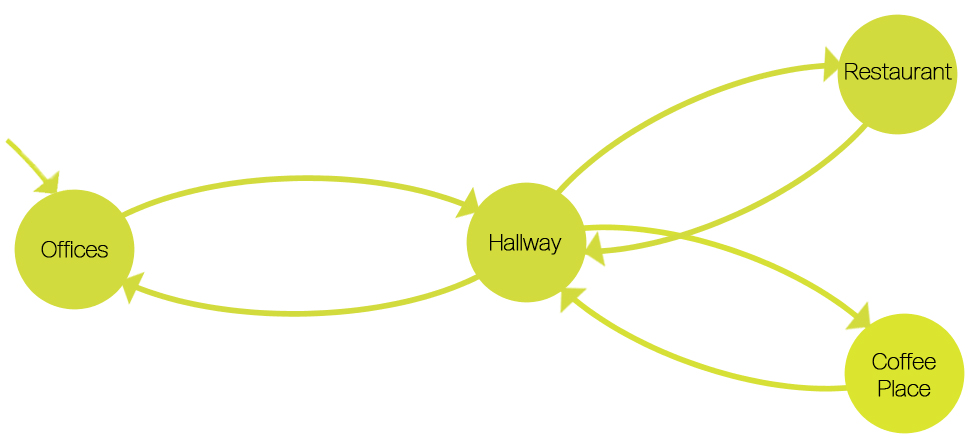}
\caption{The moving model for agents}
\label{fig:movingDiagram}
\end{figure}

The moving submodel has four components that agents can transfer between. As it is shown in Figure \ref{fig:movingDiagram} the initial state is offices and from that state agents can leave their offices and pass the hallway to either have lunch at the restaurant or grab a cup of coffee at the coffee place. During the interviews, we discovered most of the people do those daily routines around the same period of time and only a few people do not follow this pattern and leave their offices out of usual time periods, so we concluded the transition of agents between different locations is a random phenomenon which follows a normal probability distribution. 

\begin{figure}[H]
\includegraphics[width=\linewidth, center]{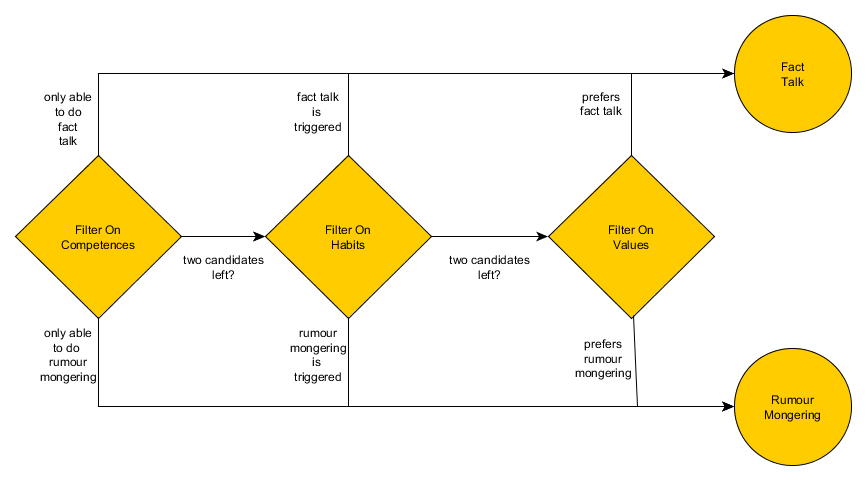}
\caption{The choose-activity submodel and the three stages the agent uses to decide on its activity: competences, habits and values.}
\label{fig:choose-activity}
\end{figure}

The choose-activity submodel is based on \citet{Mercuur2017a} and has three stages. The submodel is depicted in Figure \ref{fig:choose-activity}. The agent starts by considering both rumourmongering and fact talk. At each stage the agent makes a decision on one cognitive aspect. If this aspect is not conclusive it will prolong the decision to the next stage. In the first stage, the agent compares its own competences to the competences that it beliefs to be required for the activity. In our example model depicted in Figure \ref{fig:umlinstance}, Bob would decide it cannot do the activity of \texttt{rumourmongering}, because it requires a competence he does not have: noticing juicy details. As such, Bob will engage in fact talk. (Note that if Bob does not have the skill to do either activity, then the decision is also prolonged to the next stage.) In the second stage, an agent tries to make a decision based on its habits. It will survey its context and decide which context elements are near, i.e., resources, places or other agents. If it has a habitual trigger association with a particular strong strength between one of those context elements and either rumourmongering or fact talk it will automatically do that action. In the last stage, the agent will consider how strongly it relates certain values to both activities and how strongly it adheres itself to these values. Consequently, it makes a comparison between the two activities and decides which best suits its values. For the complete implementation of the habitual model and value model we refer to \citep{Mercuur2015}.

\section{Experiment}\label{experiments}
The proposed rumour model with elements associated with physical settings, individuals’ values and competencies enables us to investigate impacts of a variation of settings and interventions on the spread of rumours in organizations.  

One of the open questions in organizational rumour literature is the effectiveness of different prevention and control strategies. In our approach we only need to extend the model with the specific elements and characteristics of the case that we would like to study. In this paper we study the effect of organizational layout on rumour dynamics. In our case, we take the size of offices and number of coffee places as the proxies for organizational layout and juxtapose two organizational layouts cases (Figure 4) to understand the impact of layout on rumourmongering dynamic.

To setup the model, we determine the number of agents, then initialize the context and agents. In the organization that we studied each section has on average 50 people, therefore, we pick 50 as the number of the agents. For context initialization, we design the layouts and assign agents to different locations, then we initialize agents with probability distributions for routines such as grab a cup of coffee or having lunch. After the model setup, it can be executed. 

\begin{figure}[H]
\includegraphics[width=\linewidth, center]{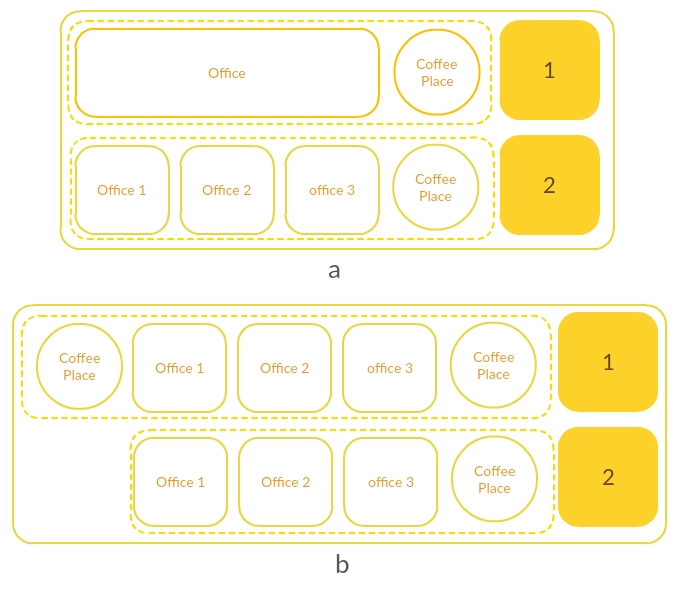}
\caption{(a) In this case, we study the impact of office size on dynamics of rumourmongering (b) In this case we study the impact of number of coffee places on the dynamics of rumourmongering}
\label{fig:lay}
\end{figure}

%
%
%

\section{Discussion \& Future Research}\label{concl}
Modelling rumourmongering has been studied since 1964. So far, the modelling did not consider the complexities of individual agents, and mostly focused on the spreading behaviour of the phenomenon. In the model proposed in this paper, agents have a cognitive layer that  deploys social practice theory and views rumour as a routine with associated competencies, values and a physical setting. 

In this research, we narrowed our study to the context of organization and after introducing the generic model, we tailored our model to the context of organization via empirical data collected though interviews conducted in a Dutch University. Based on explorative interviews we established that social practice theory are likely to be applicable as people shared a view on rumour, and their habits regarding rumour and rumours seem to be intertwined with other activities.

Our model can be used to study a wide range of topics in organizational rumour studies, in particular for testing the effectiveness of interventions for prevention and control of rumours in organizations. 

Future work is to extend the questionnaire by asking about associations, conduct more and more rigorous interviews, implement the model and run the proposed experiments that explore different organization layouts. Furthermore, we aim to validate our model by looking at how rumours travel from person to person in the organization during a pre-selected time period.
                   
\section*{Contributions \& Acknowledgements}
Fard \& Mercuur wrote the first draft. Fard provided the domain knowledge and collected most data, whereas Mercuur provided the meta-model and methodological knowledge. Dignum, Jonker and van der Walle. supervised the process and contributed to the draft by providing comments, feedback and rewriting. This research was supported by the Engineering Social Technologies for a Responsible Digital Future project at TU Delft and ETH Zurich.

\bibliographystyle{splncsnat}
\bibliography{Mendeley_Rijk-Amir_ESSA2018.bib}

\end{document}